\documentstyle[twoside,fleqn,espcrc2,epsf]{article}

\newcommand{\AmS}{{\protect\the\textfont2
  A\kern-.1667em\lower.5ex\hbox{M}\kern-.125emS}}

\def\Rin{R_{\rm in}}
\def\dof{\,{\rm dof}}
\def\Rg{R_{\rm g}}

\def\plotone#1{\centering \leavevmode
\epsfxsize=0.45 \textwidth \epsfbox{#1}}

\def\plotbig#1{\centering \leavevmode
\epsfxsize= 0.9\textwidth \epsfbox{#1}}

\def\mdcrit{{\dot m}_{\rm crit}}
\def\dm{{\dot m}}

\title{Relativistic smearing of the reflection spectrum in Galactic Black Hole Candidates}

\author{C. Done, P.T. Zycki\address{Department of Physics, University of Durham,
        South Road, Durham DH1 3LE, UK }
        and 
        D.A. Smith\address{Department of Physics and Astronomy, University of Leicester,
        University Road, Leicester, LE1 7RH}}
       
\begin{document}

\begin{abstract}

We identify the reflected component in the GINGA spectra of Nova Muscae, a
Black Hole transient system which has been used as the prototype for the
recent advection dominated disk models. We see that the reflected spectrum is
generally significantly relativistically smeared, and use this and the amount
of reflection to track the innermost extent of the accretion disk. We see
that the optically thick disk does retreat during the decline, but more
slowly than predicted by the advective models, posing problems
for this description of the accretion flow.

\end{abstract}

\maketitle

\section{INTRODUCTION}
 
Black hole binary systems give one of the most direct ways in which to study the
physics of accretion disks. There is no surface boundary layer or strong central
magnetic field to disrupt the flow, and the orbital parameters are often well
studied so that the inclination, mass and distance of the system are tightly
constrained. Additionally, many of these systems (the Soft X--ray Transients,
hereafter SXT's) show dramatic outbursts where the luminosity rises rapidly from
a very faint quiescent state to one which is close to the Eddington limit, and
then declines again over a period of months, giving a clear sequence of spectra
as a function of mass accretion rate. 
The usual pattern for such objects is for the
outburst to be dominated by a soft component of temperature $\sim 1$ keV, with
or without a (rather steep) power law tail, while the later stages of the
decline show much harder power law spectra, extending out to 100--200 keV
(see e.g. \cite{TS96}).
The same bimodal spectral states
are seen in the persistent black hole candidates (such as Cyg X--1), showing
that they are a general outcome of an accretion flow.

The ``standard'' accretion disk model developed by \cite{SS73};
hereafter SS, derives the accretion flow structure in the limit when the
gravitational energy released is radiated locally in a geometrically thin disk.
Such models give temperatures of order 1~keV for galactic black hole candidates 
(hereafter GBHC) at high accretion rates
but are unable to explain the presence of a (hard or soft) power law tail to
high energies. Either there are parts of the disk flow in a different
configuration to that of SS, or there is some non--disk structure such as a
corona powered by magnetic reconnection (e.g. \cite{HMG94}), 
for which there is little physical justification as yet.

\begin{figure*}[!ht]
\plotbig{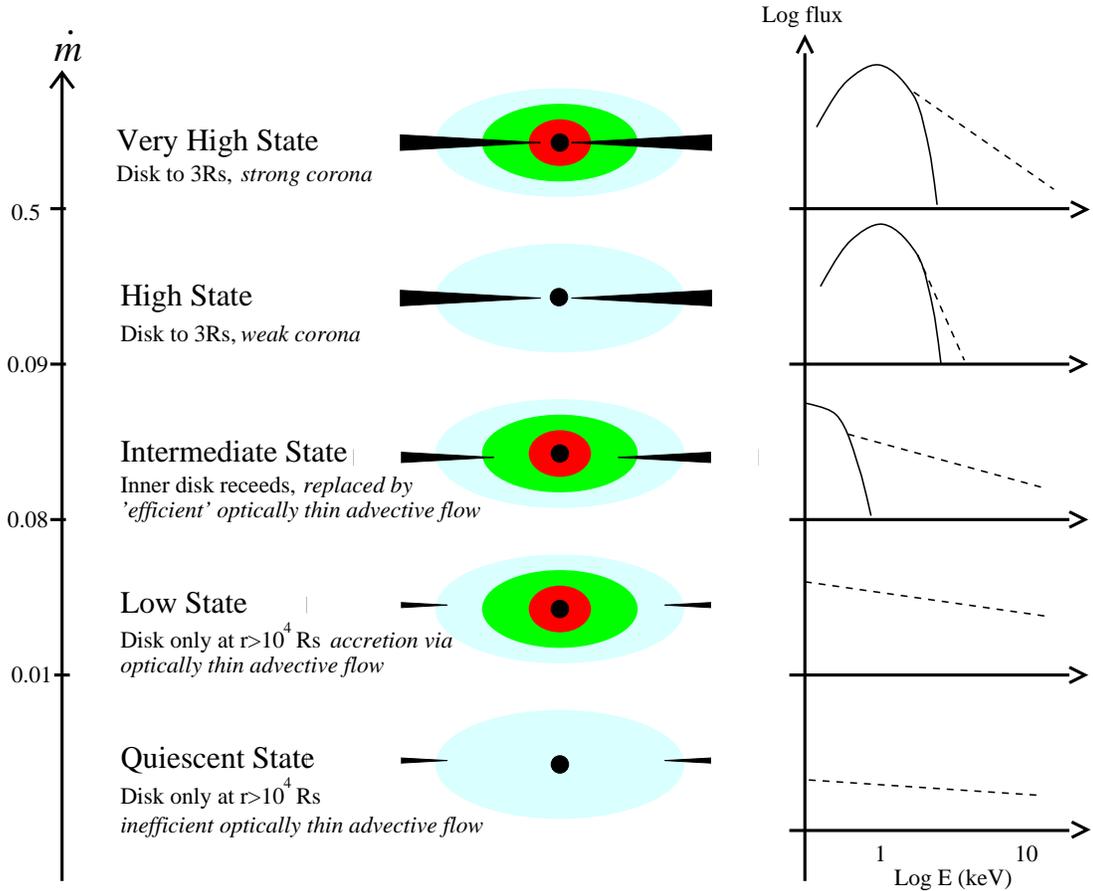}
\vskip -20pt
\caption{Schematic view of the evolution of the accretion flow in an SXT
outburst proposed by \cite{Esin97} as a function of mass accretion. Typical
spectra for each state are shown also, with the solid line and dashed lines indicating the 
SS disk and advective flow emission, respectively
}
\end{figure*}

Recently there has been much excitement about the possibility that another
solution of the accretion flow may explain the hard X--ray data without having
recourse to such {\it ad hoc} structures. Below a critical accretion rate,
$\dm\le \mdcrit$, a stable, hot, optically thin, geometrically thick solution
can be found if radial energy transport (advection) is included (see e.g.\ the
review by \cite{Nar97}). The key assumption of these models is that the protons
gain the energy from the viscous processes, while the electrons only acquire
energy through interactions (electron-ion coupling) with the protons. This
coupling timescale can be rather slow compared to the accretion timescale, so
protons can be accreted into the black hole, taking some of the 
energy with them.  The energy that the electrons do manage to obtain is radiated
away via cyclotron/synchrotron emission (on an internally generated magnetic
field), bremsstrahlung, or Compton scattering of the resultant spectra of these
two processes. In contrast to the SS accretion flow models, there is no cold
disk, so no strong source of soft seed photons for Compton scattering, hence the
resulting X--ray spectra are hard. 

Such flows were proposed to explain
the hard and very faint X--ray spectra seen from SXT in quiescence
(e.g. \cite{NMY95}), and then extended by \cite{Esin97}
to cover the whole range of luminosity seen in SXT's. As the mass
accretion rate increases from $\dm << \mdcrit$ to $\dm\sim \mdcrit$ the flow
density increases, so the electron--ion coupling becomes more
efficient and the fraction of energy the electrons can drain from the protons
increases. This increase in radiative efficiency continues to $\dm=\mdcrit$,
where only $\sim 35$ \% of the heat energy is advected. At higher $\dm >
\mdcrit$ the advective flow collapses into an SS disk. This change 
from a hot advective flow to a cool SS disk is postulated as the origin of the hard / soft 
spectral transition seen in GBHC \cite{Esin97}. Their scenario for the SXT spectral evolution
is shown in Figure 1. 

\section {TESTING THE ADVECTIVE FLOW MODELS}

\begin{figure}[ht]
\plotone{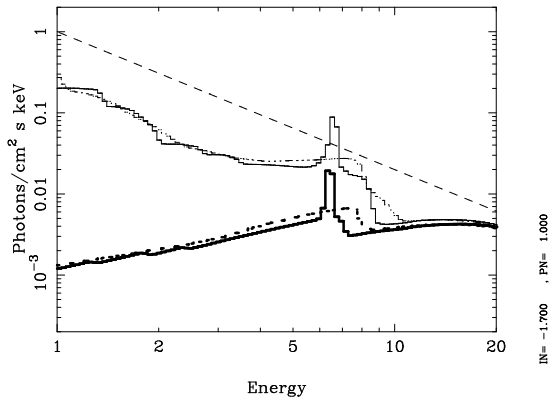}
\vskip -20pt
\caption{
Reflected continuum and associated iron line from power law illumination of
optically thick, cosmic abundance material subtending a solid angle of $2\pi$
and viewed at an angle of $60^\circ$. The thick solid line shows the reprocessed
spectrum from neutral material while the thin solid line shows the spectrum from
highly ionised material. The thick and thin dotted lines show the expected
distortions from relativistic smearing if the material forms a disk extending
from infinity to $6\Rg$, where $\Rg=GM/c^2$.}
\vskip -20pt
\end{figure}

The changing geometry shown in Figure 1 is {\it testable} from the X--ray
spectral data.  Wherever hard X--rays illuminate optically thick material, such
as a cool SS disk, there is the possibility that the X--rays can be reflected
back through electron scattering. The reflection probability is given by a
trade--off between the importance of electron scattering and photo--electric
absorption.  Higher energy photons are then preferentially reflected due to the
smaller photo--electric opacity of the material, and there is associated
fluorescence line emission from the recombining photo--ionised species. 
The dependence 
on photo--electric absorption means that the reflected spectrum
is a function of the ionisation state and elemental abundances of the
reflecting gas. 
The thick solid line in Figure 2 shows the 
characteristic 1--20 keV reflected spectral shape from neutral material, while the thin solid
line shows the resulting reflected spectrum from highly
ionised material, assuming a power law illuminating
spectrum (dashed line in figure 2) \cite{LW88,GF91,MPP91}.

The sharp spectral features shown in Figure 2 are what we expect to see from
hard X--ray illumination of a slab of material. For material around a black
hole, the conditions are somewhat different! The high orbital velocities give 
relativistic Doppler shifts, and gravitational redshift
is also important. The expected reflected line profiles are broad and skewed
rather than narrow atomic transitions \cite{Fab89}. The effect of these
relativistic processes on the total reflected spectrum from both neutral and
ionised material is shown as the thick and thin dotted lines in figure 2.  The
sharp line features are almost entirely lost, while the edge is only slightly
broadened.

These reprocessed spectral features give us a strong diagnostic of the geometry.
The amount of reflection and line indicate the solid angle subtended by the
optically thick material, while the relativistic smearing shows where this
material is in the gravitational potential. Together this means that the
structure of the accretion flow can be assessed from high quality hard X--ray
data, giving a clear test of the theoretical accretion models discussed
above. In particular, reflected features can only be seen if there is optically
thick material that subtends a substantial solid angle to the X--ray source,
unlike the advective source geometry for the low and quiescence 
state SXTs \cite{Esin97}.

We used the GINGA data on the SXT Nova Muscaee (GS 1124--68) to look for these
effects, since it is the source used by \cite{Esin97} to illustrate their
model, and shows the full sequence of spectral evolution from Very High to Low
state shown in Figure 1.  We fit these data with self--consistent reflected
continuum plus line models, including relativistic
smearing (see Figure 2 and \cite{ZDS97}). 

\begin{table*}
\setlength{\tabcolsep}{1.5pc}
\newlength{\digitwidth} \settowidth{\digitwidth}{\rm 0}
\catcode`?=\active \def?{\kern\digitwidth}
\begin{tabular*}{\textwidth}{@{}l@{\extracolsep{\fill}}lccccc}
\hline

 data  & $\Gamma$ & $f$ &
$\xi$   & $\Rin/\Rg$ &
$\chi^2/\dof$ \\

\hline

Jan 11, VHS   & $ 2.03\pm 0.16$           &    $0.30^{+0.13}_{-0.04} $  &
         $(1.0^{+1.0}_{-0.7})\times 10^4 $ &
         $13^{+5}_{-3}$        & 26.1/31\\
May 18, HS/IS & $ 2.29^{+0.05}_{-0.03} $  &    $ 0.64^{+0.40}_{-0.10} $  &
         $ (3.5^{+9.5}_{-3.0})\times 10^4$ & $18^{+22}_{-6} $ & $13.5/22 $ \\
June 13, IS & $1.91\pm 0.03$  &  $0.57_{-0.17}^{+0.23}$  &
        $10^{+22}_{-8} $  &  $10^{+9}_{-4}$ &      $20.9/24$ \\
June 21, IS & $1.83\pm 0.03$ &  $0.46\pm 0.12$  &
        $ 4^{+12}_{-3}$ & $13^{+25}_{-6}$    &       $11.9/24$ \\
July 23, LS &  $1.72\pm 0.02 $  &  $0.24_{-0.08}^{+0.11} $  &
        $17^{+40}_{-16} $  &  $50^{+\infty}_{-35}$   & $15.3/24 $ \\
Sept 3, LS  & $1.95\pm 0.11 $   &  $0^{+1.1}$ &
     0(f)  & -- & 26.9/26 \\

\hline
\end{tabular*}
\end{table*}

In the Very High State the reflector is highly ionised and strongly
relativistically smeared. The power law component is too weak to examine
for reflected features in the
High State, but at the start of the Intermediate State the reflector is still
highly ionised and strongly relativistically smeared. It seems probable that
during all of this time the
disk extends down to the last stable orbit, and that it is ionised by the strong
soft component. As source fades through the Intermediate and Low states, the
importance of the soft component declines, the ionisation of the reflector
drops, the power law hardens by $\Delta\Gamma\sim 0.4$, and the solid angle $f$
subtended by the reflector decreases (see table 1).

\section{CONCLUSIONS}

All the observed spectral changes during the Intermediate and Low State 
can be explained in a model
where the inner radius of the optically thick flow progressively increases: this
would trivially explain the decrease in 
solid angle subtended by the disk to the inner X--ray source. Also, as the
luminosity and temperature of the disk declines (since most of the energy is
released in the inner radii) so the ionisation state of the disk decreases,
and there are fewer seed photons for Compton scattering, giving a harder power law.

Thus, the observed transition from high (soft) to low (hard) state does seem to
involve a retreat of the optically thick material, as proposed by \cite{Esin97}.
However, quantitatively, the inner disk radius we infer is much smaller than
that of their models for the intermediate and low states \cite{Esin97}. 
Their concept of accretion
occuring via an optically thin flow from very large radii ($10^4 Rs$) for the
low state spectra cannot be sustained as such models produce negligible
reflected features.  Instead we see reflection at a level at a level which is
lower than expected from a complete disk, but is nonetheless
significant. Similar covering fractions for the reflector are seen in other GBHC
hard state spectra, showing that it is a general property of these sources
(\cite{Done92,UED94,Ebi96,Gier97,Dove97,PKR97}).  
While the calculation of 
the transition radius between the advective and SS flow
is rather ad hoc,  nonetheless a major
facet of their model is the sudden switching of {\it all} the accretion flow
into an optically thin state over a very small range in $\dot{m}$ \cite{Esin97}. This is
clearly inconsistent with the observed behaviour of GBHC in general and Nova
Muscae in particular, where a composite optically thin/optically thick flow is
required.

\end{document}